\documentclass[sigconf,authorversion=true,nonacm=true]{acmart}
\usepackage{enumitem}

\usepackage{booktabs} 

\usepackage{csquotes}
\usepackage{amsmath}
\usepackage{bm}
\usepackage{outlines}

\usepackage{etoolbox}
  \robustify\bfseries

\usepackage{siunitx}
\usepackage{graphicx}

\usepackage{pifont}
\usepackage{array}
\newcolumntype{P}[1]{>{\centering\arraybackslash}m{#1}}
\newcolumntype{L}[1]{>{\arraybackslash}m{#1}}

\newcommand{\dd}{\mathop{}\!\mathrm{d}}

\usepackage{xspace}
\newcommand\ie{i.\,e.\xspace}
\newcommand\eg{e.\,g.\xspace}

\usepackage{url}

\newcommand{\mcellt}[2][c]{%
	\begin{tabular}[t]{@{}#1@{}}#2\end{tabular}}

\makeatletter
\renewcommand{\fps@figure}{htb}         
\renewcommand{\fps@table}{htb}         
\makeatother 

\usepackage{bbm}
\usepackage[ruled]{algorithm2e}

\setcopyright{acmcopyright}
\copyrightyear{2022}
\acmYear{2022}
\acmDOI{10.1145/3485447.3512000}

\acmConference[WWW '22]{Lyon '22: ACM The Web Conf}{April 25--29, 2022}{Lyon, France}
\acmBooktitle{Proceedings of the ACM The Web Conf (WWW '22), April 25--29, 2022, Lyon, France}
\acmPrice{}
\acmISBN{}

\newcommand{\model}{MMHM\xspace}

\hypersetup{draft}

\begin{document}
\fancyhead{}

\title[Detecting False Rumors from Retweet Dynamics on Social Media]{Detecting False Rumors from Retweet Dynamics on Social Media}


\author{Christof Naumzik}
\affiliation{%
	\institution{ETH Zurich}
	\city{Zurich}
	\state{Switzerland}
}
\email{cnaumzik@ethz.ch}

\author{Stefan Feuerriegel}
\affiliation{%
	\institution{LMU Munich}
	\city{Munich}
	\state{Germany}
}
\email{feuerriegel@lmu.de}

%
\renewcommand{\shortauthors}{Naumzik and Feuerriegel}

\begin{abstract}
False rumors are known to have detrimental effects on society. To prevent the spread of false rumors, social media platforms such as Twitter must detect them early. In this work, we develop a novel probabilistic mixture model that classifies true vs. false rumors based on the underlying spreading process. Specifically, our model is the first to formalize the self-exciting nature of true vs. false retweeting processes. This results in a novel mixture marked Hawkes model (\model). Owing to this, our model obviates the need for feature engineering; instead, it directly models the spreading process in order to make inferences of whether online rumors are incorrect. Our evaluation is based on 13,650 retweet cascades of both true. vs. false rumors from Twitter. Our model recognizes false rumors with a balanced accuracy of \SI{64.97}{\percent} and an AUC of \SI{69.46}{\percent}. It outperforms state-of-the-art baselines (both neural and feature engineering) by a considerable margin but while being fully interpretable. Our work has direct implications for practitioners: it leverages the spreading process as an implicit quality signal and, based on it, detects false content. 
\end{abstract}

%
%

\ccsdesc[500]{Information systems~Social networks}
\ccsdesc[500]{Mathematics of computing~Stochastic processes}
\ccsdesc[100]{Information systems~Data mining}

\keywords{Spreading process; Hawkes process; Bayesian modeling; Mixture model; Fake news; Social media}

\maketitle

\section{Introduction}
\label{sec:introduction}

Social media platforms have emerged as a prevalent information source for large parts of the society, thereby introducing an unprecedented change to contemporary news consumption \cite{Bakshy.2015}. It is currently estimated that almost 62~percent of the adult population rely upon social media platforms for this purpose and this proportion is expected to increase further \cite{Pew.2016}. At the same time, social media often lack proper quality control over the content and are thus especially prone to false rumors.

False rumors refer to online content that disseminates \textquote{fabricated information that mimics news media content in form but not in organizational process or intent} \cite{Lazer.2018}. False rumors are thus intentionally and verifiably wrong (see \cite{Vosoughi.2018} for a definition). An example of a false rumor is, for instance, the so-called Pizzagate that erroneously alleged Hillary Clinton of running a child sex ring \cite{Kim.2019}. False rumors have evolved into a widespread concern for society. According to recent surveys, the number of social media interactions with news stories labeled as false has surpassed the number of engagements with mainstream media \cite{TheEconomist.2017}. Moreover, the average US adult was estimated to have seen and remembered \num{1.14} incorrect news stories prior to the same election, which, at least for some, has affected their decision making \citep{Allcott.2017}.  Hence, there is a direct need for social media platforms to better detect and eventually mitigate false rumors.    

Detecting false rumors on social media must fulfill several \textbf{requirements from practice} (\eg, \cite{Lazer.2018,TheEconomist.2017,Liu.2018}): 
\begin{enumerate}[leftmargin=0.5cm]
\item A corresponding detection mechanism must respond in a timely manner in order to provide early warnings. As prominent examples, the Cyberspace Administration of China has recently imposed severe penalties for hosting fake news \citep{WSJ.2017} and German law even requires hate speech and fake news to be deleted within 24 hours \citep{TheEconomist.2017}. Because of this, the possibility for human fact-checking by domain experts is restricted. Currently, such checks are subject to a severe time lag, oftentimes surpassing 24 hours \citep{Shao.2016}. As a remedy, the use of statistical classifiers for pre-filtering which tweets are sent to manual fact-checking seems beneficial. 
\item A detection mechanism must be scalable. The reason is that contemporary social media platforms receive more than 500,000 tweets per minute that must be processed \cite{TheEconomist.2017}. Given the sheer data volume, not only the sensitivity of a statistical classifier is of importance but also false alarms can be problematic, since each incurs substantial follow-up costs due to manual fact-checking. Accordingly, a statistical classifier should be cost-effective in the sense that, for a high detection rate, the specificity should also be favorable. 
 
\item A detection mechanism must be dynamic. That is, it should not be trained towards certain topics that have been in the training set and are thus a~priori known (\eg, climate change) but it should also recognize rumors about new topic that were previously unseen (\eg, rumors related to the COVID-19 pandemic). Therefore, it was argued earlier that, in practice, a detection mechanism should not rely upon individual linguistic cues \cite{Liu.2018} but should consider only the general topic of a tweet. A classifier should rather rely on user covariates and, in particular, retweet dynamics. As such, the data structure of retweet cascades must be modeled, that is, the branching which gives rise to self-exciting processes \cite{Crane.2008}. 
\end{enumerate}

\noindent
Several statistical classifiers have been developed to detect false rumors (cf. Sec.~\ref{sec:background}). However the above requirements from practice demand that only features from users and retweet dynamics are considered. In keeping with this, there have been attempts to test whether structural features of cascades could potentially lend to prognostic capacity: on the one hand, this idea has been validated based on feature engineering \cite{Castillo.2011,Kwon.2013,Kwon.2017}. Here a limited set of hand-engineered features was crafted, whereby cascades are mapped onto aggregated variables. However, due to the aggregation, a large extent of the predictive power is lost. On the other hand, prior works have developed classifiers that take the sequence of retweets as input (\eg, \cite{Liu.2018,Ma.2016,Vosoughi.2015}). As such, the temporal dimension of retweets is maintained. However, this stream has overlooked a unique property of retweet cascades \cite{Crane.2008}: retweet dynamics are self-exciting (due to the branching inside the cascade). This gap is filled by our work.

\vspace{0.15cm}
\noindent \textbf{Objective:} Can we model the different self-exciting spreading processes of true vs. false rumors on social media platforms in order to predict the veracity of tweets?
\vspace{0.15cm}

To achieve this objective, research must follow a theory-informed approach and encode theoretical properties of online spreading. Specifically, online spreading processes are \textbf{self-exciting} (\eg, \cite{Crane.2008,Mishra.2016,Zhao.2015}). That is, the occurrence of past tweets makes sharing and thus the occurrence of future retweets more probable. Yet the self-exciting nature has not yet been leveraged for predicting whether rumors are true vs. false. Here our work is based on the hypothesis that the true vs. false rumors are characterized by different self-exciting processes that can be used for prediction. By modeling the self-exciting nature of online spreading, our work gives a rise to several benefits: we consider the re-sharing dynamics in spreading process and, following this, each individual retweet acts as a form of quality control.

\textbf{Novelty:} Our work aids social media platforms such as Twitter as follows. We present the first statistical model that separates the spreading processes of true vs. false rumors. For this, we formalize the self-exciting nature of online spreading processes via a novel \emph{mixture marked Hawkes model}~(\model). Our proposed model accommodates various variables that describe the heterogeneity in the spreading process at both tweet and user level. We evaluate our statistical model for detecting false rumors on the basis of 13,650 retweet cascades from Twitter. Our proposed model recognizes false rumors with a balanced accuracy of \SI{64.97}{\percent} and an AUC of \SI{69.46}{\percent}. We compare our proposed model against various baselines -- both neural and with feature engineering -- that represent the state-of-the-art. These cannot capture the self-exciting nature of retweet dynamics (and thus the branching inside cascades), whereas, in our model, its is considered via a mixture of two marked Hawkes processes. As a result, the baselines are outperformed by our \model in terms of AUC by a considerable margin but while being fully interpretable. Code and data made publicly available at \url{https://github.com/cfnaumzik/FakeNewsDetection}.

\textbf{Practical significance:} This research was conducted in way that it closely matches the needs from industry practice such as at Twitter:
\begin{enumerate}[leftmargin=0.5cm]
\item \emph{Early warnings.} Our model is highly effective as an early warning system. To achieve this, we opted for a parsimonious model specification, which yields a superior performance in data-sparse settings, \ie, when only a few retweets are available for detecting false rumors. This is confirmed in our computational experiments, where the AUC remains robust at around 0.70, independent of whether a prediction is made 24~hours or 30~min after the initial tweet. 
\item \emph{Cost-effectiveness.} Our model is on par with state-of-the-art baselines in terms of sensitivity but, in addition, achieves a specificity is that is substantially better. As such, our model reduces the cost for false alarm (\ie, due to manual fact-checking) by a significant margin.   
\item \emph{Crowd intelligence.} Our proposed model leverages crowd behavior as a warning mechanism. It is based on the hypothesis from social sciences stipulating that false rumors travel different than true rumors \cite{Vosoughi.2018}. As such, we use retweet dynamics as an implicit quality control and thus make predictions based on them. This ensures a straightforward application in practice.  
\end{enumerate}
\noindent
To facilitate industry uptake, both code and data are made publicly available at \url{https://github.com/cfnaumzik/FakeNewsDetection}.

\section{Background}
\label{sec:background}

\subsection{Information Spreading in Social Media}

Social media allows for users to passes content (\ie, tweets) to their follower base \citep[\eg,][]{Myers.2012}. This thus gives rise to distinctive statistical properties describing online spreading \citep[\eg,][]{Cha.2009,Goel.2012,Goel.2015,Kwak.2010,Lerman.2010,Leskovec.2007,Macskassy.2011,Myers.2014,prollochs2021emotions,prollochs2021emotions2}. Specifically, if many re-shares occur, the source tweet can go {viral} and receive considerably popularity. \textbf{self-exciting} process \citep[cf.][]{Crane.2008,Kobayashi.2016,Mishra.2016,Zhao.2015}. The self-exciting nature motivates our model development later. 

Online spreading has been used for making various inferences. Examples of corresponding outcomes are the overall reach of a cascade \citep[\eg,][]{Kupavskii.2012,Myers.2012,Subbian.2017,Weng.2013,chen2019information} or its relative growth \citep[\eg,][]{Cheng.2014,Yang.2010}. Other classification labels are the veracity of rumors or whether a retweet cascade belongs to a rumor vs. fact; see our elaborations later. 

A common approach for modeling retweet cascades is to directly model the self-exciting nature of spreading processes via a (marked) Hawkes process (\eg, \cite{Kobayashi.2016,Mishra.2016,Zhao.2015}). Thereby, the inherent mechanism of re-sharing is translated into a parsimonious formalization, and, as result, it reflects empirical evidence that the arrival rate of retweets is non-constant \cite{Crane.2008}. Hawkes processes are usually used to forecast the expected lifetime of a cascade, whereas our objective is to classify the spreading process itself (and thus the rumor veracity). Because of this difference, the aforementioned references assume a \emph{single} spreading process where all retweet cascades are formalized via the \emph{same} set of parameters. However, our task is based on \emph{different} spreading processes for true vs. false rumors that needs to be classified. The latter requires a different model (\ie, a mixture) which presents the added contribution of our work. 

\subsection{Rumor Spreading in Social Media}

Descriptive studies have advanced our understanding of rumor spreading in social media. These works have compared structural characteristics of retweet cascades from rumors vs. facts (\ie, rumors vs. non-rumors). This includes various structural properties of retweet cascades for instance, their lifetime \cite{Castillo.2011,prollochs2021emotions,prollochs2021emotions2}, number of retweets \cite{Friggeri.2014,prollochs2021emotions,prollochs2021emotions2}, and emotions \cite{Zeng.2016}. We are aware of only one work that explicitly compares statistical properties of true vs. false rumors \cite{Vosoughi.2018}. Here summary statistics suggest that false rumors coincide -- on average -- with a faster, deeper, and broader spread.

Preventing false rumors has received considerable traction. Fact-checking was studied earlier, suggesting that it is effective \citep{Tambuscio.2015} but oftentimes lags the original dissemination by 24 hours \citep{Shao.2016}. A remedy is seen in statistical classifiers. For a detailed survey, we refer to \citep{Kumar.2018,Zubiaga.2018} and only summarize key concepts in the following. We emphasize that many of the following works classify rumors vs. facts (\ie, rumors vs. non-rumors), whereas we are interested in the more challenging task of discerning true vs. false rumors.

Predictions based on rumor cascades can be loosely grouped in two paradigms: (i)~\emph{Feature engineering} is used to summarize the shape of retweet cascades, so that a traditional classifier (\eg, random forest) can then be used. Examples of the underlying features include the out-degrees, number of nodes, depth-to-breadth ratio, or retweeting time (cf. \cite{Castillo.2011,Castillo.2013,Kwon.2013,Kwon.2017,Vosoughi.2017}). (ii)~A second paradigm builds upon sequential models, where the cascade is mapped on time series (longitudinal) structures. In this regard, examples of sequential learning comprise hidden Markov models \cite{Vosoughi.2015,Vosoughi.2017} and long short-term memory networks \cite{Ducci.2020,Ma.2016}. Alternatively, pruning is applied to preprocess the variable-length sequence, so that gated recurrent units \cite{Ma.2016} and fused variants with convolution neural networks \cite{Liu.2018} can be used. These present later our baselines. However, the self-exciting nature of retweeting is ignored in the aforementioned works and is unique to our model. 

\subsection{Self-Exciting Point Processes}

A popular self-exciting process is the so-called {Hawkes process} \citep{Hawkes.1971}, for which the intensity function depends on all previous events. It describes processes where one event is likely to trigger subsequent events (such as in the case of retweeting) \citep{Crane.2008}. Relevant to our work is a specific variant, \ie, the marked Hawkes process. It carries an additional mark that captures information on the event. Marked Hawkes processes have seen a wide range of application in data mining, \eg, in modeling earthquakes \citep{Ogata.1988} or location traces \citep{du2016recurrent}. While mixtures of self-exciting process have been developed \cite{Yang.2013}, we are not aware of applications involving mixtures of \emph{marked} Hawkes process.

Due to their self-exciting nature, Hawkes processes have been used to model retweet cascades in social media (\eg, \cite{Kobayashi.2016,Kong.2020,Medvedev.2019,Mishra.2016,Zhao.2015}). Here the findings provide consistent evidence that marked Hawkes processes are highly effective for modeling the spreading behavior of online information. In this regard, Hawkes processes model the self-exciting nature of retweet cascades, so that the branching inside a retweet casacade is implicitly considered. Specifically, it was shown to be linked to the susceptible-infections-response process in epidemic modeling \citep{Kong.2020}. Note that, in previous works, the marks are simply the number of followers \cite{Kobayashi.2016,Mishra.2016,Zhao.2015}, whereas the marks in our \model are a linear model that incorporates additional covariates as another source of heterogeneity. 

However, these models are not suitable as baselines. Importantly, these works \cite{Kobayashi.2016,Kong.2020,Medvedev.2019,Mishra.2016,Zhao.2015,chen2019information} are concerned with the prediction of the final size of a retweet cascade and not rumor veracity. That is, they can only learn parameters of a \emph{single} spreading process (\ie, an infectiousness parameter for predicting the final size) and thus cannot distinguish among \emph{multiple} spreading processes. Formally, the underlying prediction frameworks make predictions \emph{within} a given cascade and cannot make classifications \emph{between} cascades. As such, the aforementioned works are single Hawkes processes \cite{Kobayashi.2016,Kong.2020,Medvedev.2019,Mishra.2016,Zhao.2015,chen2019information,du2016recurrent}, whereas, in contrast, we develop a mixture model. Because of this, their application as baselines is precluded. 

\textbf{Research gap:} To the best of our knowledge, a self-exciting process for classifying the spreading process of online rumors into true vs. false has not yet been developed. While mixtures of self-exciting process have been developed \citep{Yang.2013}, we are not aware of applications involving mixtures of \emph{marked} Hawkes process.  To close this gap, we model the self-exciting spreading processes of true vs. false rumors via a novel mixture marked Hawkes model.

\section{The Proposed Mixture Marked Hawkes Model (\model)}
\label{sec:model}

\subsection{Overview}

In our work, the objective is to determine the veracity $V(\mathcal{C})$ that belongs to a retweet cascade $\mathcal{C}$. For this, the proposed \model models the actual spreading process behind $\mathcal{C}$, which is defined by the following input and output. 

\textbf{Input:} The \model is fed with a retweet cascade $\mathcal C= \left(\bm z,\bm{\tau}\right)$, which comprises two sets of variables: the covariates $\bm z\in\mathbb{R}^{n_c}$ describe the original source tweet (\eg, the topic) and the longitudinal variable $\bm{\tau} = \left(\bm{\tau}_i\right)_{i=0}^n$ which encode all retweets $i = 1, \ldots, n$ of the source tweet. In terms of notation, an index $i = 0$ refers to the initial tweet of a cascade (\ie, its root / source), while $i > 0$ refers to a \emph{re}tweet thereof. As such, the variable $\bm{\tau}$ is responsible for specifying the complete retweet structure.

Each (re)tweet is associated with further covariates, \ie, $ \bm{\tau}_i = \left(t_i,p_i,\bm x_i,\bm y_i\right)$, as follows: $t_i$ with $t_0\leq t_1\leq\ldots\leq t_n$ denotes the time of a (re)tweet $i$. For simplicity, we set $t_0 = 0$. The variable $\bm{x}_i\in\mathbb{R}^{n_u}$ refers to additional covariates at user level (\eg, the number of followers). The variable $\bm{y}_i\in\mathbb{R}^{n_s}$ accommodates variables that describe structural properties of a cascade (\eg, the current depth, response time for a retweet). Finally, the variable $p_i$ specifies the tree structure underlying a cascade, that is, it points towards the parent tweet, \ie, $p_i\in\{0,\ldots,i-1\}$. Note that the original source tweet entails two sources of heterogeneity: the variable $\bm z$ defines general properties that reflect the topic, whereas $\bm{\tau}_0$ describe the dynamics that occur with all tweets (\eg, the number of followers or other variables at user level).

\textbf{Output:} The output is a binary label $V(\mathcal{C})$ describing whether a spreading process stems from a tweet that is true ($=0$) or declared as false ($=1$). We explicitly encode false rumors as $=1$ since false rumors should trigger an alarm. By using this encoding, the precision directly measures how accurately false rumors are detected.\footnote{True positive: candidate of a false rumor that turned out to be actually false; False positive: candidates of a false rumor that turn out to be actually true; etc.} At this point, we acknowledge that the veracity of retweet cascades is not necessarily binary, as oftentimes content is of mixed veracity. However, fake news detection in practice demands a binary classification as actions of social media platforms are binary (\ie, whether to show a warning or not). This thus gives our inference task, namely $P(V(\mathcal{C}) \,\mid\, \bm z,\bm{\tau})$. 

\subsection{Model Components}

Our proposed \model combines five components as follows. A \textbf{(C1)~mixture} is responsible for encoding the different spreading behavior of true vs. false rumors. Each spreading behavior is formalized as a tailored point process, \ie, a \textbf{(C2)~marked Hawkes process}. It utilizes a \textbf{(C3)~marked memory kernel}, which allows us to accommodate heterogeneity in the excitement of tweets and time dynamics. Both function in slightly different ways, and, for that reason, we specify a model for the \textbf{(C4)~marks} and a \textbf{(C5)~baseline kernel}, respectively.

\vspace{0.2cm}
\noindent
\textbf{Mixture (C1): }
The \model makes predicts the veracity $V(\mathcal{C})$ via a mixture. That is, we fit two separate processes $M_F$ and $M_T$ to retweet cascades that are either true or false, respectively. Both models have different parameters and thus capture how true vs. false rumors travel. When applied to unseen retweet cascades, we determine which set of model parameters -- \ie, $M_F$ or $M_T$ -- is more suitable in describing the underlying spreading process. Both $M_F$ and $M_T$ are specified as a marked Hawkes model in the following. For ease of notation, we omit the subscript and simplify write $M_{\ast}$ when referring to the underlying marked Hawkes process for both veracity labels $\ast \in \{ F, T \}$. 

\vspace{0.2cm}
\noindent
\textbf{Marked Hawkes process (C2):} 
We follow \cite{Kobayashi.2016,Mishra.2016,Zhao.2015,hatt2020early} and model the longitudinal data from the retweet cascades as a point process, specifically a Hawkes process. For this, we first review point processes. These provide a natural way to describe random points (\ie, retweets or other events) over a continuous time period. This is followed by a marked Hawkes process $M_{\ast}$ that is tailored to rumor spreading.

\begin{figure}[htbp]
\centering
\vspace{-0.3cm}
\includegraphics[width=.9\linewidth]{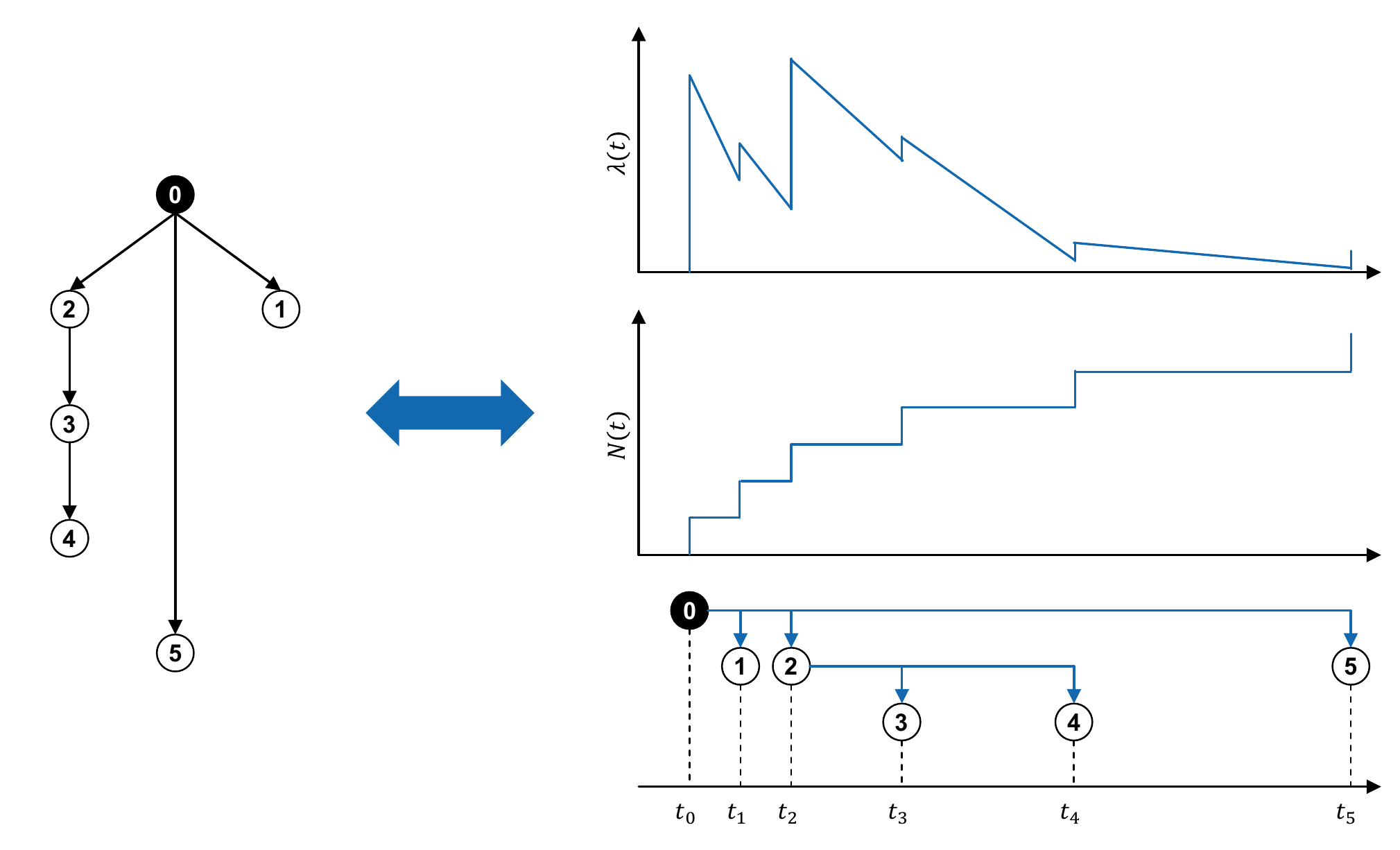}
\vspace{-0.5cm}
\caption{Example of a retweet cascade (left) and its corresponding point process (right).}
\label{fig:cascadepointprocess}
\end{figure}

A {point process} is uniquely characterized by a counting measure $N(t)$ \cite{VereJones.2003}. In our case, it is simply the size of the retweet cascade (\ie, the number of retweets) up to time $t$, \ie,
\begin{equation}
N(t) = \sum_{i=0}^n\mathbf{1}_{\{t_i\leq t\}}.
\end{equation}
Then, the so-called intensity function $\lambda(t)$ gives the probability of observing a retweet in an infinitesimal small interval at time $t$, \ie,
\begin{equation}
\lambda(t) = \lambda(t \mid \mathcal{H}_t) = \lim\limits_{h\to0}\frac{N(t+h) - N(t)}{h} ,
\end{equation}
where $\mathcal H_t$ denotes prior tweet history up to time $t$, that is, the sequence $t_1,\ldots,t_{N(t)}$. We omit $\mathcal H_t$ for ease of notation.  Fig.~\ref{fig:cascadepointprocess} provides an example, in which a retweet cascade is translated into a point process. By formalizing the structure of $\lambda(t)$, we can incorporate further heterogeneity and time dynamics. 

We model $\lambda(t)$ so that we yield a marked Hawkes process. Thereby, we combine two characteristics. First, as any Hawkes process \cite{Hawkes.1971}, the intensity function $\lambda(t)$ depends on all retweets (\ie, events) that occurred up until time $t$. This thus yields a process that is self-exciting in nature. Second, the marks allow us to further cater for heterogeneity in the the influence of past tweets (\ie, some can have more excitation than others). For this, each tweet $i$ carries in addition to the time $t_i$ also a mark $m_i\in \mathbb{R}^{+}$ that models that influence on retweeting. As we shall see later, it accommodate various covariates that describe the overall excitation of a tweet. Hence, it thus reflects heterogeneity in the propensity that a tweet will be shared (\eg, some users might have a larger follower base and thus a larger reach). 

Formally, the intensity function $\lambda(t)$ is modeled via
\begin{equation}
\label{eq:hawkes_intensity_short}
\lambda(t) = \sum_{i \colon t_i<t}\phi_{m_i}(t-t_i) ,
\end{equation}
where $\phi_{m_i}(s)\colon\mathbb{R}_+\to\mathbb{R}_+$ is the so-called \emph{marked memory kernel}. The marked memory kernel $\phi_{m_i}(t-t_i)$ modulates the effect of the event $i$ on the intensity $\lambda(t)$. As such, it is responsible for modeling the time dynamics. For instance, if we choose $\phi_{m_i}(t)$ to be a monotonically decreasing function\footnote{We note that this choice is not per~se needed during estimation, but it reflects a common intuition and is consistent with prior work on information cascades \cite{Kong.2020}.}, more recent retweets will have a larger influence on the current intensity rate. In addition, the shape of the memory kernel determines how fast a retweet $i$ is ``{forgotten}'' and thus reflects the duration during which the rumor is active and thus likely to be re-shared. In this sense, the model implicitly considers that each re-share (\ie, event) is triggered by a specific previous event, either the original message or another re-share.

The above formalization incorporates prior theory on rumor spreading. First, the arrival rate of new retweets is not constant \cite{Wu.2007}, and, hence, $\lambda(t)$ is modeled as a function of time. Second, the arrival rate of future retweets depends on already observed tweets (\ie, by summing over $i \colon t_i<t$). As a result, the arrival rate should increase when more users are exposed to a tweet, thus reinforcing the chance of the retweet cascade to grow further. This is reflected in the self-exciting nature of the Hawkes process. Third, heterogeneity is captured by the mark $m_i$.

\vspace{0.2cm}
\noindent
\textbf{Marked memory kernel (C3): }
 Following \cite{Kobayashi.2016,Mishra.2016,Zhao.2015}, we model the marked memory kernel $\phi_{m_i}$ as the product of mark $m_i\in\mathbb R_+$ and a baseline memory kernel $\phi_0\colon\mathbb{R}_+\to\mathbb{R}_+$, \ie,
\begin{equation}
\phi_{m_i}(s) = m_i\,\phi_0(s) .
\end{equation}
This allows us to separately model two effects: On the one hand, we capture the strength of a specific (re)tweet on the intensity function and thus future reweets via the mark $m_i$. It also controls for various covariates that describe how engaging users and the content of the source tweet is. On the other hand, we model the duration of the effect via the baseline memory kernel $\phi_0$. It specifies the time dynamics with which the tweet influences the intensity function and thus future retweets. 

\vspace{0.2cm}
\noindent
\textbf{Mark model (C4): }
 The mark $m_i$ for tweet $i$ should further capture the different sources of heterogeneity in online information diffusion. These stem from cascade ($\bm{z}$), user ($\bm{x}$), and structural ($\bm{y}$) covariates. Therefore, we model the mark $m_i$ as a log-linear function of different covariates, \ie, 
\begin{equation}
\label{eq:mark_linear_model}
\log\left(m_i\right) = \alpha + \beta_c^T\bm z_i + \beta_u^T\bm x_i + \beta_s^T\bm y_i
\end{equation}
with intercept $\alpha$ and coefficients $\beta_c$, $\beta_u$, and $\beta_s$ for cascade, user, and structural covariates, respectively. Our particular choice of covariates is detailed in Sec.~\ref{sec:covariates}.  The linear formalization helps us in obtaining a parsimonious model. This is desirable for two reasons. First, it reduces the risk of overfitting and, second, it is interpretable. Accordingly, we can directly quantify the contribution of individual covariates to the number of future retweets. 

The above formalization allows to incorporate prior theory on the rumor spreading. (1)~The mark, and thus implicitly the intensity function, accounts for heterogeneity in the content of the source tweet via $\bm{z}$. For instance, some content is simply more engaging than others \cite{Mishra.2016}. (2)~There exists differences between users regarding their social influence. For example, users with more followers tend to create more re-shares \cite{Kobayashi.2016,Mishra.2016,Zhao.2015}. Accordingly, the propensity of retweeting should depend on the social influence $\bm{x}$. (3)~The arrival rate of retweets over time is not constant \cite{Wu.2007}. For instance, structural information, such as the elapsed time since the original message was created, also influences the number of re-shares. Hence, the mark, and thus implicitly $\lambda(t)$, must be a function of time. This is achieved by setting one of the covariates in $\bm{y}$ to the elapsed time. 

The above formalization reveals a crisp difference to prior models. The latter simply set the mark $m_i$ to the number of followers \cite{Kobayashi.2016,Mishra.2016,Zhao.2015}. In contrast, we incorporate further sources of heterogeneity by formalizing $m_i$ as a linear model.

\vspace{0.2cm}
\noindent
\textbf{Baseline memory kernel (C5): }
 The baseline memory kernel $\phi_0$ models how the influence of a tweet on the intensity function decays over time. Hence, it implicitly quantifies the time dynamics of excitation. In our model, we accommodate further flexibility by allowing for different baseline memory kernels at the source tweet (\ie, the root of retweet cascade) and all \emph{re}tweets (\ie, all non-roots). We thus yield two baseline memory kernels, namely $\phi_0^\text{root}$ and $\phi_0^\text{non-root}$, respectively. This is motivated by empirical observations according to which the distribution of retweet times varies significantly between source tweets at the root and subsequent retweets. 

Our implementation draws upon a power law kernel for root messages and an Weibull kernel for non-root messages. This allows us to reflect that the duration until a human response to a tweet follows a heavy-tailed distribution. As part of our robustness checks, we later confirm that this choice is superior.  

\subsection{Inferring Rumor Veracity from Spreading Processes}

As detailed earlier, the mixture is responsible for estimating two different spreading processes for true vs. false rumors. This yields two separate marked Hawkes models with different parameters $M_T$ and $M_F$, respectively. We then aim to infer the veracity for an unseen retweet cascade $\mathcal{C}^{\ast}=\left(\bm{z}^{\ast},\bm{\tau}^{\ast}\right)$. The inference is formalized via the following classification task.

First, we evaluate the likelihood under the model $M_T$ and the model $M_F$ for the given data $\mathcal{C}^{\ast}$. This yields $P\left( \bm{z}^{\ast},\bm{\tau}^{\ast}\mid M_T \right)$ and $P\left( \bm{z}^{\ast},\bm{\tau}^{\ast}\mid M_F \right)$. Second, we compute the conditional probability of $\mathcal{C}^{\ast}$ belonging to $M_F$, \ie, $P\left( V(\mathcal{C}^{\ast}) = 0 \mid \bm{z}^{\ast},\bm{\tau}^{\ast} \right)$. It is obtained by applying Bayes' theorem via
\begin{equation}
\label{eqn:bayes_inferences}
P\left( V(\mathcal{C}^{\ast}) = 0 \mid \bm{z}^{\ast},\bm{\tau}^{\ast} \right) = \frac{P\left( \bm{z}^{\ast},\bm{\tau}^{\ast}\mid M_F \right)}{P\left( \bm{z}^{\ast},\bm{\tau}^{\ast}\mid M_F \right) + P\left( \bm{z}^{\ast},\bm{\tau}^{\ast}\mid M_T \right)} .
\end{equation}
Here all terms on the right-hand side can be directly inferred from the estimated mixture model. Third, the probability is mapped onto a binary label analogous to other classification tasks. This thus infers the veracity of a tweet based on its spreading behavior.

\subsection{Model Estimation}

The proposed \model is estimated based on full Markov chain Monte Carlo~(MCMC) sampling \cite{Gelman.2014} by directly sampling from the posterior distribution of the model parameters. For this, we derived the likelihood function $\mathcal{L}$ (see Appendix~\ref{appendix:estimation_details}). We checked the fitted model by following common procedures for point processes, \ie, goodness-of-fit checks \cite{Clements.2012} and posterior predictive checks \cite{Gelman.2014}. Both checks were positive (see Appendix~\ref{appendix:checks_model_fit}).  

We emphasize a particular strengths of our \model. In contrast to conventional machine learning, hyperparameters are absent. Therefore, the need for parameter tuning is circumvented. Instead, all parameters can be directly estimated from data. 

\section{Setting}
\label{sec:setting}

\subsection{Data}

The automated detection of rumors (\ie, isolating them from non-rumors) can nowadays be achieved with very high, almost perfect accuracy \cite{Ma.2015,Ma.2017,Liu.2018}. However, the subsequent step of identifying its veracity (\ie, classifying whether such a previously-identified rumor is true or false) is considerably more challenging and presents thus focus of our work. 

In this work, we use the dataset from \cite{Vosoughi.2018}, which comprises actual retweet cascades from rumors on Twitter. Permission from Twitter for using this dataset for the purpose of our study was obtained. The data provides a real-world, cross-sectional sample with considerable heterogeneity. This choice ensures a realistic dataset and that thus demands by social media platforms in practice are met. In fact, the dataset is widely regarding as ``representative'' and because of which the risk of potential bias should be minimal. 

The dataset was subject to a lightweight preprocessing that one would normally apply in practice; see Appendix~\ref{appendix:preprocessing_details}. The veracity label $V(\mathcal{C})$ was determined based on six independent fact-checking organizations as in \cite{Vosoughi.2018}. The majority vote over the fact-checking votes is taken. Note that the fact-checking organizations have an almost perfect pairwise agreement (Fleiss kappa of above 95\,\%). The resulting dataset comprises 817,131 individual tweets from 13,650 retweet cascades (with a class imbalance of 84.34\,\% vs. 15.66\,\%). The class imbalance was addressed as follows: During model estimation, we randomly selected \num{750} cascades for each veracity. This thus forms our training set. The remaining cascades represent our test set (\ie, {12,150} cascades).

\subsection{Model Variables}
\label{sec:covariates}

Our \model model is fed with the following variables. We remind that our model parameterization is flexible and, if used at a social media platform other than Twitter, can be updated in a straightforward manner (in fact, all variables are also available at other platforms like Facebook, WeChat, Parler, etc.). 

\textbf{Cascade covariates ($=\bm{z}$):} First, emotions have been identified in social psychology research as a key driver behind consumption \cite{Upworthy} and rumor diffusion \cite{Rosnow.1991}. We measured the emotions embedded by different tweets based on the NRC dictionary \cite{Mohammad.2013}. As suggested earlier \cite{Kusen.2017}, we use a low-dimensional representation: (i)~\emph{positive} emotions refer anticipation, joy, and trust; (ii)~\emph{negative} emotions includes anger, disgust, fear, and sadness; and (iii)~\emph{surprise} was kept as a separate category.

Second, the topic of a source tweet might be linked to the propensity of retweeting. This hypothesis was confirmed for tweets on political content \citep{Vosoughi.2018}. Hence, we include a dummy variable that indicates whether the message covers a political subject.

\textbf{User covariates ($=\bm{x}$):} Each tweet $\bm{\tau}_i$ is associated with variables characterizing the social influence of a user. Following \cite{Vosoughi.2018}, we collect (i)~the \emph{number of followers} of the user; (ii)~the \emph{number of accounts followed}; (iii)~a user's \emph{account age}; and (iv)~her \emph{engagement}. The latter is the past activity on Twitter measured as the total number of tweets, re-tweets, replies, and likes relative to the account age. Altogether, these variables control for potential between-user heterogeneity in activity patterns. Note that we refrain from using actual user histories as this would restrict inferences to users that have been active (and thus prevents cold-start settings). 
 
\textbf{Structural covariates ($=\bm{y}$):} These variables encode dynamic properties of the spreading process, \ie, they leverage the retweet cascade until the current time $t_i$. We include (i)~the \emph{depth} of $\bm{\tau}_i$, that is, the number of events on the direct path between $\bm{\tau}_i$ and $\bm{\tau}_0$ in the cascade; (ii)~the total \emph{elapsed time} since event $\bm{\tau}_0$, \ie, $t_i-t_0$; and (iii)~the \emph{response time} $t_i-t_{p_i}$. These variables are potentially informative for two reasons. First, these variables control for the previously stated observation that the so-called {infectivity}, that is, the number of retweets of a tweet, decays over time \cite{Kobayashi.2016,Wu.2007}. Second, prior research \cite{Vosoughi.2018} has identified significant differences in the diffusion speed, as well as the depth of the information cascades belonging to true vs. false rumors.

\subsection{Summary Statistics}

The average number of tweets per retweet cascade amounts to \num{59.86}. It is slightly larger for false than for true rumors. On average, the number of leaves in a cascade (\ie terminal retweets) amounts to \num{56.67}. 

This sheds light into the branching structure behind spreading dynamics. When diving the number of retweets by the number of followers for a user, we find that, on average, a tweet is shared merely by \SI{0.27}{\percent} of a user's followers. The average lifetime of a retweet cascade spans \num{33.62} hours, during which it can influence public opinions negatively. A tweet is shared within a median time frame of \num{102.73} minutes. This duration is shorter for false rumors than for true ones. This is in line with prior evidence suggesting that false rumors travels faster, wider, and deeper \cite{Vosoughi.2018}. Summary statistics of the model variables ($\bm{x}$ and $\bm{y}$ as logs) are reported in Tbl.~\ref{tbl:model_variables}.

\begin{table}[H]
\centering
\sisetup{parse-numbers=false}
{\scriptsize
\begin{tabular}{l cc}
\toprule
Covariate & False rumor & True rumor \\
\midrule
\multicolumn{3}{l}{{Cascade covariates} $=\bm{z}$} \\
\midrule
Positive emotion &  0.3324 & 0.3462 \\
Negative emotion & 0.5210 & 0.5211 \\
Surprise emotion & 0.1449 & 0.1292 \\
Topic dummy ($=1$ if political content) & 0.5669 & 0.5983 \\
\midrule			
User covariates $=\bm{x}$  \\
\midrule			
Number of followers & 6.2744 & 6.2621 \\
Number of accounts followed & 6.3768 & 6.3831 \\
Account age (in days) & 6.7566 & 6.8515 \\
Engagement (user interactions / account age) & 2.4229 & 2.4580 \\
\midrule
Structural covariates $=\bm{y}$ \\
\midrule
Depth & 1.4390 & 1.4154 \\				
Response time (in hours) & 1.3105 & 1.2372 \\
Elapsed time (in hours) & 1.5476 & 1.4511 \\
\bottomrule	
\multicolumn{3}{l}{Values in $\bm{x}$ and $\bm{y}$ as logs; additional interaction terms as detailed later}
\end{tabular}
}
\footnotesize
\caption{Model covariates and summary statistics (mean).}
\label{tbl:model_variables}
\end{table}

\subsection{Baselines}

Our proposed \model model is compared against several state-of-the-art approaches for classifying retweet cascades. These follow two different paradigms: (1)~sequential models where the spreading process is mapped onto a time series and where thus the self-exciting nature of the retweet patterns is lost, and (2)~feature engineering on top of traditional machine learning where the aggregated features are computed, yet where the informativeness of individual retweet events are lost. Details are in Appendix~\ref{appendix:baselines}.

\textbf{Sequential models:} These models have access to individual retweets as our \model:
\begin{enumerate}[leftmargin=0.5cm]
	\item {Time-series SVM (\textbf{ts-SVM})} \cite{Ma.2015}: This maps the cascade onto a dynamic series-time structure.  
	\item {Gated recurrent units (\textbf{GRU})} \cite{Ma.2016}: This work applies a pruning step to the variable-length sequential data, so that it can then be fed into a GRU.
	\item {Long short-term memory (\textbf{LSTM})} \cite{Ma.2016}: This forms a specific recurrent neural network where the architecture is designed to process long-raging sequences. 
	\item {Fused GRU-CNN network (\textbf{GRU-CNN})} \cite{Liu.2018}: This model prunes the sequential data and then combines a gated recurrent unit and a convolutional neural network for the purpose of rumor detection. 
	\item Plain \model \cite{Mishra.2016}: This model is inspired by previous research on formalizing retweet cascades via marked Hawkes processes \citep{Kobayashi.2016,Mishra.2016,Zhao.2015}. However, these models cannot be directly applied as a baseline; instead, they must be extended by a mixture to discriminate two different spreading processes analogous to component~(C1) of our \model. In the plain \model, the linear model for the mark from Eq.~(\ref{eq:mark_linear_model}) is simplified analogous to \cite{Kobayashi.2016,Zhao.2015} by setting the mark $m_i$ to the number of followers, yet we follow \cite{Mishra.2016} and include all user covariates $\bm x$.  
\end{enumerate}

\noindent
The above models all introduce a fairly large parameter space and thus interpretability is largely precluded. Hence, we also opt for \textbf{feature engineering} as a common approach from the literature to obtain parsimonious model formulations similar to our \model. Our hypothesis is that such a parsimonious model might be effective in data-sparse setting, especially in the first time window shortly after the original source tweet. Here we closely adhere to prior literature \cite{Castillo.2011,Conti.2017,Kwon.2013,Vosoughi.2017}, which leads to the following theory-informed choice of aggregated features: size, depth (overall and average), size-to-depth ratio, average response time, average elapsed time, structural virality, diffusion speed, average follower number, average account age, average followee number, average user engagement, topic dummy, emotions (negative, positive, surprise). To this end, these feature cover both spreading dynamics and user covarates similar to \model, yet the between-tweet variance is largely lost during aggregation. In contrast, this information is explicitly kept in our proposed model. 

\textbf{Additional baselines} (omitted for brevity). (1)~We use a HMM with multivariate Gaussian emission probability as in \cite{Vosoughi.2015,Vosoughi.2017} but its AUC was only 58.02 and thus lower than all other baselines. (2)~We also experimented with the convolutional network in \cite{chen2019information} and tree-structured LSTM \cite{Ducci.2020} but found that both are inferior and, thus, were omitted. This can be explained by the fact that both have large parameter spaces that may not be effective in complex yet data-scarce settings. (3)~We also adopt a ``single Hawkes process'' approach, in which we first fit Hawkes process to a given rumor cascade and, in a second step, use the estimated parameters for predicting the veracity. This is in line with \cite{Kobayashi.2016,Kong.2020,Medvedev.2019,Mishra.2016,Zhao.2015,chen2019information,du2016recurrent}. However, this approach was consistently inferior. In general, our inferences via Bayes' theorem (see Eq.~\ref{eqn:bayes_inferences}) follow the same logic. However, our mixture does not rely on point estimates when comparing the likelihood of parameters for true/false rumors under a rumor cascade; instead, we use a fully Bayesian approach and thus account for the complete posterior distribution. Hence, explaining why our mixture approach for making predictions is superior. 

\section{Empirical Evaluation}
\label{sec:evaluation}

\subsection{Estimated Model Parameters}

Tbl.~\ref{tbl:estimated_parameters} lists the estimation results for the two mixture components in the \model. The dependent variable is the mark, $\log{m_i}$, and, hence, it quantifies the diffusion intensity, where a larger value is associated with more retweets. As part of our estimation procedure, the variables were transformed in a way (cf. Appendix~\ref{appendix:estimation_details}) that it measures the expected number of retweets. Reiteratingly, the coefficients measure the contribution of a variable ceteris paribus, that is, in keeping all other covariates equal. This allows us to determine how the spreading process links to the veracity of rumors. 

Evidently, true rumors attract more retweets than false ones as demonstrated by a larger intercept. Social influence is linked to more retweets for false rumors than for true rumors. This becomes evident by larger coefficients for the number of followers and the number of accounts followed. In keeping everything else equal, social influence has a stronger effect on the overall reach of a cascade for fake rumors. A different finding is made for engagement: more engagement with false rumors diminishes their spreading, whereas it gives more retweets for true rumors. However, false rumors travel wider due to the fact that depth is not reducing the retweeting to the same extent as it does for true rumors. The effect for emotion is more substantial for false rumors (both for positive and negative), yet, interestingly, not for ``surpise'' as an affective dimension. Finally, political tweets capture more interest and have a higher propensity to be retweeted in the case of false rumors than for true rumors.

\begin{table}[H]
	\centering
	{\scriptsize
	  \sisetup{parse-numbers=true,detect-weight}
		\begin{tabular}{ll S[table-format=2.1,round-precision=2,round-mode=places]
				S[table-format=2.1,round-precision=2,round-mode=places]}
			\toprule
			& Variable & {\mcellt{False rumors ($M_F$)}} & {\mcellt{True rumors ($M_T$)}} \\			
			\midrule
			$\alpha$ & Intercept & \bfseries 1.6447 & \bfseries 2.3440\\[0.1cm]
			$\bm{x}$ & Number of followers & \bfseries 0.2047 & \bfseries  0.1959 \\
			& Number of accounts followed & \bfseries 0.3996 & \bfseries 0.1166 \\
			& Account age & -0.0215 & -0.0269  \\
			& Engagement &\bfseries -0.0785 &  \bfseries 0.3738  \\[0.1cm]
			$\bm{y}$ & Depth & \bfseries -2.8421 & \bfseries-3.2102 \\
			& Response time & -0.1004 & -0.0921 \\
			& Elapsed time & -0.3302 &-0.3207 \\[0.1cm]
			$\bm{z}$ & Positive emotions &\bfseries0.2983 & \bfseries 0.1594  \\
			& Negative emotions & \bfseries-0.4507 &\bfseries -0.1721 \\
			& Surprise &\bfseries0.0255 & \bfseries 1.9251  \\
			& Topic & \bfseries0.2160 &\bfseries-0.1983 \\[0.1cm]
			\bottomrule
		\end{tabular}
	}
	\footnotesize
	\caption{Estimated spreading process parameters in the \model for true vs. false rumors Parameters which are significantly different from each other at the \SI{0.01}{\percent} level are highlighted in bold.}
	\label{tbl:estimated_parameters}
\end{table}

\subsection{Out-of-Sample Accuracy in Detecting False Rumors}

Tbl.~\ref{tbl:discrimanatory_power} examines the out-of-sample ability of the models to detect false rumors based on the complete retweet cascade. Recall that the label $V(\mathcal{C})=1$ correspond to a false rumor. The best-performing model is given by our \model with an AUC of 69.46. Compared to the best baseline (fused GRU-CNN \cite{Liu.2018}), this is an improvement by \num{3.07} percentage points. The improvement is statistical significant at the 0.1\,\% level. For the F1 score, the improvement of our \model over the second best model even amounts to 12.77\,\%. Also, our \model is the best with regard to precision.

\begin{table}[H]
\centering
{\scriptsize
\begin{tabular}{l SSSS SS}
\toprule
{Model} & {AUC} & {\mcellt{Balanced\\ accuracy}} & {Sensitivity} & {Specificity} & {Precision} & {F1}  \\
\midrule
\multicolumn{5}{l}{\emph{Sequence learning}} \\
\midrule
ts-SVM \citep{Ma.2015} & 60.81  & 57.90 &  61.86 & 53.94 & 14.75 & 23.83 \\	
GRU \citep{Ma.2016}&60.40  &52.68 &47.08 &64.17 & 14.48 & 22.15 \\
LSTM \citep{Ma.2016}&60.57&52.55&62.65 &49.95 & 13.89 & 22.74 \\
Fused GRU-CNN \citep{Liu.2018} & 66.39 & 60.89 & 56.95 & 65.60& 17.58 & 26.87 \\
Plain \model \citep{Mishra.2016} & 58.40 & 59.55 & 62.22 & 56.88 & 15.68 & 25.05 \\			
\midrule
\multicolumn{5}{l}{\emph{Feature engineering} \citep[cf.][]{Castillo.2011,Conti.2017,Kwon.2013,Vosoughi.2017}} \\
\midrule
Logistic regression & 60.47 & 58.36 & 58.39 & 58.31 & 15.29 & 24.23 \\
Random forest & 64.45 & 59.96 & 59.34 & 60.59 & 16.25 & 25.51 \\	
Gradient boosted m. & 63.07 & 59.13 & 61.86 & 56.39 & 15.45 & 24.73 \\
Na{\"i}ve Bayes & 57.26 & 55.74 & 59.69 & 51.79 & 13.76 & 22.36 \\
Neural network & 60.47 & 58.36 & 58.39 & 58.31 & 15.29 & 24.23 \\
\midrule
Proposed \model & \bfseries 69.46  & \bfseries 64.97 &  61.28 & \bfseries 68.66 & \bfseries  20.13 & \bfseries  30.30\\			
\bottomrule	
\multicolumn{5}{l}{Stated: metric in \%; best value highlighted in bold}
\end{tabular}
}
\footnotesize
\caption{Out-of-sample accuracy in detecting rumor veracity.}
\label{tbl:discrimanatory_power}
\end{table}

Our \model outperforms state-of-the-art sequence learning models but while being fully interpretable. For comparison, the second best baseline is a complex Fused GRU-CNN for which interpretability is precluded. Given the complexity of the best state-of-the-art baseline in comparison to the parsimonious structure of our proposed MHMM, we think that the performance improvement is remarkable. The gain can be attributed to the fact that the \model models the self-exciting structure whereas the baselines consider retweet cascades as flat sequences without branching structure. Notably, our \model is largely on par with other models in terms of sensitivity but, at the same time, the specificity is substantially improved: for sensitivity of around 61\,\%, the \model achieves a specificity of 68.66\,\%, which, compared to other baselines, is a plus by at least 3 percentage points This is desired due to practical requirements, as false alarms incur substantial downstream costs (\eg, for manual fact-checking and the threat that many false alarms lead users ignoring warning messages on social media websites). 

Note that some caution is needed when comparing our detection accuracy to prior literature (\eg, \cite{Ma.2015,Ma.2016,Liu.2018}). Therein, the objective was different (\ie, to discriminate rumors vs. non-rumors), whereas we engage in the more challenging task of classifying true vs. false rumors. 

\subsection{Early Warnings: Detection Accuracy for Partial Retweet Cascades}

In a next step, we assess the ability of our model to act as an early warning system. Hence, we measure the performance for partial retweet cascades: (i)~after a certain amount of elapsed time (Tbl.~\ref{tbl:discrimanatory_power_time}) and (ii)~after a certain number of retweets (Tbl.~\ref{tbl:discrimanatory_power_events}). As a disclaimer, some models have a larger AUC than before, which is attributed to the complex model structure, that can lead to overfitting on larger cascades. In comparison, our \model can successfully make inferences even in data-scarce settings and is thus more effective than state-of-the-art baselines on both partial and complete retweet cascades.

To ensure a fair comparison, all of the benchmark models were re-estimated for each variation in time and cascade size. When inferring the veracity of a rumor 30\,min after the source tweet, the \model achieves an AUC of 70.22. The prediction performance of the baselines is outperforemd and remains fairly stable independent on the size of the time window.\footnote{We considered all baselines, but only show the two best options from each category for reasons of brevity.} This suggests that the spreading process provides a powerful predictor of veracity. Similar findings are obtained when making predictions after a small number of retweets. Here only 5 retweets seem sufficient to yield an AUC of \num{70.97}. These findings highlight the benefits of our model specification: the \model is parsimonious and, even in data-scarce settings, highly effective.

\begin{table}[h]
\centering
{\scriptsize
\begin{tabular}{l SSSSSSS}
\toprule
&\multicolumn{7}{c}{Time window (in hours)}\\
\cmidrule(lr){2-8}
{Model} & {0.5} & {1} & {2} & {6} & {12} & {24} & {168} \\
\midrule
ts-SVM \citep{Ma.2015} & 60.17 & 61.30 & 61.72& 61.43& 61.18& 60.68 &60.81\\
Fused GRU-CNN \citep{Liu.2018} & 66.08 & 68.79  & 67.00  & 67.47 & \bfseries 71.50 & 69.11 & 66.02 \\
Random forest & 62.87 & 63.96& 63.10 &64.07 &63.99& 63.86& 64.45 \\
Gradient boosted machine & 60.01 & 61.81& 61.76 &63.17 &62.55&  62.57& 62.89\\
\midrule
Proposed \model & \bfseries 70.22 & \bfseries 70.11 & \bfseries 69.93 & \bfseries69.67 & 69.32 & \bfseries69.21 &\bfseries 69.46 \\
\bottomrule	
\end{tabular}
}
\footnotesize
\caption{Performance (AUC) of early detection after a certain lifetime.}
\label{tbl:discrimanatory_power_time}
\end{table}

\begin{table}[h]
\centering
{\scriptsize
\begin{tabular}{l SSSSSSS}
\toprule
&\multicolumn{6}{c}{Number of observed retweets}\\
\cmidrule(lr){2-8}
{Model} & {5} & {10} & {25} & {50} & {100} & {250} & {500} \\
\midrule
ts-SVM \citep{Ma.2015}&61.78 &61.37& 60.81& 60.54 &60.31& 60.22& 60.34 \\
Fused GRU-CNN \citep{Liu.2018} & 63.51 & 66.35 & \bfseries 72.69 & \bfseries 70.75 & 66.14 & \bfseries 71.77 & \bfseries 71.56  \\
Random forest & 64.61& 63.39& 64.89& 64.49& 64.75& 64.53& 64.77 \\
Gradient boosted machine &  63.39 & 62.79 &63.54& 62.77& 62.68& 61.94& 63.42 \\
\midrule
Proposed \model & \bfseries 70.97 & \bfseries  71.17 &  71.21 &  70.56 & \bfseries  69.87 &  69.33&  69.42 \\
\bottomrule	
\end{tabular}
}
\footnotesize
\caption{Performance (AUC) of early detection after a certain number of retweets.}
\label{tbl:discrimanatory_power_events}
\end{table}

\section{Discussion}
\label{sec:discussion}

Our proposed \model successfully manages in detecting the veracity of rumors solely from retweet processes. To achieve this, we developed a new mixture marked Hawkes model. The \model follows a theory-informed approach by modeling the self-exciting nature behind retweeting. Our \model advances prior literature in two directions: (1)~we accommodate separate spreading processes for true vs. false rumors via a mixture, and (2)~we include covariates from the between-rumor heterogeneity as part of the marks. This explains why state-of-the-art baselines are outperformed: while the baselines have access to the same data as our \model, our model is the first that accounts for the self-exciting nature of retweeting. 

We opted for a binary classification into true vs. false rumors. We are well aware that rumors can also be of mixed veracity. However, upon careful discussion with industry stakeholders, we were informed that a binary label is to be preferred. The reason is that social media platforms also follow a binary logic in their decision-making: Should a tweet cascade be subject to manual fact-checking -- yes or not? As such, our \model thus returns also a binary label that triggers an alarm or not. (We also repeated our experiments with mixed cascades that we either assigned to the true or false class, but found that our \model can still identify false rumors with higher accuracy than the baselines. However, we omitted this for brevity.)

Our model specification fulfills key requirements in practice. First, it enables early warnings in data-sparse settings. This is demonstrated in our analyses, where as few as 5 retweets are sufficient to reach an AUC of \num{70.97}. Second, our \model is designed for large-scale applications at Twitter where more than 500,000 tweets per minute must be processed. Because of this, specificity is also of practical importance (due to costs from downstream tasks such as manual fact-checking after triggering alarms). For every 500,000 tweets, the number of tweets sent to manual fact-checking at Twitter due to our model is reduced by 15,000. Third, we make only use of data from the spreading process as such data are readily available. For the same reason, data from user histories or the network structure were discarded, since, in cold-start settings, the predictive power of such data would be limited. We ensured that all covariates should also be available on other platforms (\eg, Facebook, Reddit, or Weibo). Finally, we note that the parsimonious structure of our \model allows for direct \textbf{interpretability}. Moreover, hyperparameter tuning is absent; instead, all parameters can be directly learned from data.

\clearpage

\bibliographystyle{ACM-Reference-Format-no-doi}
  \newcommand{\dq}{"}
\bibliography{literature}

\clearpage
\appendix

\section{Preprocessing}
\label{appendix:preprocessing_details}

We applied two pre-processing steps. (1)~We selected only tweets for which the fact-checking labels from the different organizations were consistently rated as \textquote{true} or \textquote{false}, excluding items with content where the veracity is potentially debatable. (2)~We removed small cascades. Here we enforced a minimum size of 6, \ie, the original tweets must have been retweeted at least 5 times. The intuition is that small cascades have less reach and are thus not as harmful as larger ones. 

\subsection{Estimation Details}
\label{appendix:estimation_details}

\underline{\textbf{Derivation of \model likelihood:}} In order to fit the Hawkes model to the data, we need to derive the log-likelihood for a given cascade with observed event times $t_1,t_2\ldots,t_n$ together with associated marks $m_1,m_2\ldots,m_n$. In practice, we only observe an information cascade during a specific time interval $\left[0,T\right]$,\eg, within the first week or the first hour. Then the log-likelihood for a marked Hawkes process \citep{VereJones.2003} is calculated as 
\begin{align}
\label{eq:ll_basic}
\mathcal{L}\left(t_1, \ldots, t_n\right) & = \sum_{i=1}^{n}\log\left(\lambda\left(t_i\right)\right) - \int_0^T\lambda\left(s\right) \dd s \\
&=
\sum_{i=1}^{n} \log\left(\sum_{j\colon t_j<t_i} m_j\, \phi_0(t_i-t_j)\right) \nonumber\\
& \hphantom{=} \quad - \sum_{i=0}^{n}m_i \, \Phi_0(T-t_i) , 
\end{align}
where $\Phi_0$ denotes the indefinite integral of $\phi_0$. 

Evaluating the log-likelihood from Eq.~\ref{eq:ll_basic} is of complexity $\mathcal{O}\left(n^2\right)$, and, hence, the estimation of a full MCMC approach becomes challenging for multiple large cascades. The complexity arises through the double summation in the first term. If we choose $\phi_0$ as the exponential kernel, a recursive evaluation scheme exists that reduces the complexity to $\mathcal{O}\left(n\right)$. However, consistent with \citep{Mishra.2016}, we find in our analyses that the exponential kernel is not flexible enough to model the intensity decay. A remedy is developed in the following.

We build upon \citep{Rasmussen.2013} and exploit the observed branching structure to simplify Eq.~\ref{eq:ll_basic}. Denoting the parent of event $i$ as $p_i$, we may write the log-likelihood as
\begin{equation}
\label{eq:ll_simple}
\mathcal{L}\left(t_1,\ldots,t_n\right) = \sum_{i=1}^{n}\log\left(m_{p_i} \, \phi_0(t_i-t_{p_i})\right) - \sum_{i=0}^{n} m_i \, \Phi_0(T-t_i).
\end{equation}
This simplification due to the explicit incorporation of the ancestral structure of the cascade introduces additional flexibility, as it renders the computation of the log-likelihood feasible for arbitrary memory kernels $\phi_0$. 

\underline{\textbf{MCMC sampling:}} Based on the derived log-likelihood, the model is estimated  via Markov chain Monte Carlo~(MCMC) sampling from the posterior. More precisely, the Hamiltonian Monte Carlo sampler (with improvement through the No-U-Turn technique) is used \citep{Hoffman.2014}, as implemented in Stan \citep{Carpenter.2017}. 

We ran two Markov chains, each with 1,000 iterations as warm-up phase and 3,000 additional iterations, yielding a total of 4,000 posterior samples for each model parameter. Owing to the efficiency of the Hamiltonian Monte Carlo sampler, a comparably small number of samples was sufficient.

The results were checked against best-practice guidelines \citep{Gelman.2014}. Specifically, the convergence of the chains was validated through checking (i)~trace plots, (ii)~the Gelman-Rubin convergence index (with a threshold of \num{1.02} for all parameters), (iii)~the so-called effective number of samples, and (iv)~whether parameters can be successfully retrieved from simulated data. All of the checks yielded positive outcomes.

\underline{\textbf{Prior choices:}} We chose weakly informative priors. Specifically, we chose standard normal priors for the parameters $\beta_c,\beta_u$, and $\beta_s$. For the mark intercept $\alpha$, we chose a Gaussian prior with standard deviation equal to $5$. Finally, for the kernel parameter, we chose standard Cauchy distributions as prior.\footnote{For a standard Cauchy distribution, the location equals $0$ and the scale parameter equals $1$.} For the Weibull kernel, we specifically restricted the shape parameter to be in $[0,1]$ to ensure that the kernel is monotonically decreasing. We achieved this by transforming the raw variable via inverse logit. We further placed a standard normal prior on the raw variable.

\section{Checks of Model Fit}
\label{appendix:checks_model_fit}

The in-sample fit of the \model to the data was checked in two ways. (1)~We ran a series of statistical tests \citep{Clements.2012} that are common in the literature for residual analysis of point processes to determine the goodness-of-fit. (2)~We conducted a series of posterior predictive checks (see \citep{Gelman.2014} for an introduction). Both are detailed in the following. Based on them, we conclude that the \model model provides a reasonable fit. 

\underline{\textbf{Goodness-of-fit checks: }}
There exist a number of statistical tests that assess the goodness-of-fit, \ie, how well point processes fit data. These perform a residual analysis to check whether the residuals are homogenous.

Formally, these tests rely upon the change time theorem \citep{VereJones.2003}. Here one tests whether the transformed (re)tweet times 
\begin{equation} 
\tau_i' = \int_0^{t_i}\hat{\lambda}(s) \dd s
\end{equation} 
follow a homogeneous Poisson process with rate 1. If that is the case, then the estimated intensity function $\hat{\lambda}$ is considered to be correct. However, the resulting tests typically have low power and are thus not suitable for our dataset. 

As an alternative, we draw upon the transformation proposed in \citep{Clements.2012} called {super thinning}. It applies a transformation that superimposes a simulated Poisson process onto the observed cascade. The joint process is then a homogeneous Poisson process of known rate if the fitted value of $\hat{\lambda}$ is correct.

In our implementation, we choose $k$ for a given cascade such that 
\begin{equation}
\inf_{t\in[0,T]}\{\hat{\lambda}(t)\}\leq k\leq \sup_{t\in[0,T]}\{\hat{\lambda}(t)\} .
\end{equation}
Then \textquote{thinning} is performed for an observed cascade by retaining each tweet $i$ with probability $p_i = \min\{ k  / \hat{\lambda}(t_i) , 1 \}$. {In our experiments, we simply set $k$ to the mean between the maximum and minimum of $\hat{\lambda}$ over the observer horizon $[0,T]$.} Based on this input, we simulate a inhomogeneous Poisson process with rate $\mu(t) = \max\{k-\hat{\lambda}(t),0\}$. The joint process, \ie, the observed cascade together with the simulated one, is then a homogeneous Poisson process with rate $k$ if the estimated intensity function $\hat{\lambda}$ is correct. Afterwards, the joint process is subject to a series of statistical tests:
\begin{enumerate}[leftmargin=0.5cm]
\item \emph{Conditional uniformity.} We check for conditional uniformity of the retweet times, that is, if $t_1/t_M,\ldots,t_{M-1}/t_M$ follow a uniform distribution on $[0,1]$. 
\item \emph{Exponential inter-arrival times.} We test if the inter-arrival times $s_i=t_i-t_{i-1}$ are exponentially distributed with rate $k$. In both cases, we apply a Kolmogoroff-Smirnov~(KS) and a Cramer-von-Mises~(CvM) test. 
\item \emph{Independence.} We use the Ljung-Box test to check for independence of the retweet times in the joint process.
\end{enumerate}
The results of all tests are show in Tbl.~\ref{tbl:gof_test}. Evidently, we cannot reject the null hypothesis at reasonable significance level for the vast majority of cascades in our training set. This in turn implies that our model manages successfully in describing the data. This is especially notable when considering that the proposed \model has only a few degrees of freedom and is thus fairly parsimonious.

\begin{table}[H]
	\centering
	{\scriptsize
		\sisetup{parse-numbers=true,detect-weight}
		\begin{tabular}{l SS SS}
			\toprule
			& \multicolumn{2}{c}{$p$-value $>0.01$} & \multicolumn{2}{c}{$p$-value $>0.05$} \\
			\cmidrule(lr){2-3}			\cmidrule(lr){4-5}
			Test & {KS test} & {CvM test} & {KS test} & {CvM test}\\\midrule
			(1) Conditional uniformity & 80.03 & 80.04 & 71.51 & 71.78 \\
			(2) Exponential inter-arrival times & 79.62 & 78.42 & 72.12 & 70.51 \\\midrule
			(3) Independence~(Ljung-Box test) & \multicolumn{2}{S}{89.72} & \multicolumn{2}{S}{84.08}\\\bottomrule
		\end{tabular}
	}
	\footnotesize
	\caption{Goodness-of-fit confirming that the proposed \model model is highly effective in describing the characteristics of data (\ie, training set).}
	\label{tbl:gof_test}
\end{table}

\underline{\textbf{Posterior predictive checks:}} Posterior predictive checks are run to gauge whether a given model has the ability to generate data that are comparable to the observed data \citep{Gelman.2014}. Specifically, we test whether retweet cascades that were simulated from our \model model display similar structures as real cascades. For this, different structural properties of spreading processes are compared, namely the (1)~size, (2)~depth, (3)~structural virality, and (4)~size-to-depth ratio. Here we refer to \citep{Goel.2012,Goel.2015} for a definition of structural virality. 

\begin{figure}[h]
\centering
\vspace{-0.5cm}
\includegraphics[width=\linewidth]{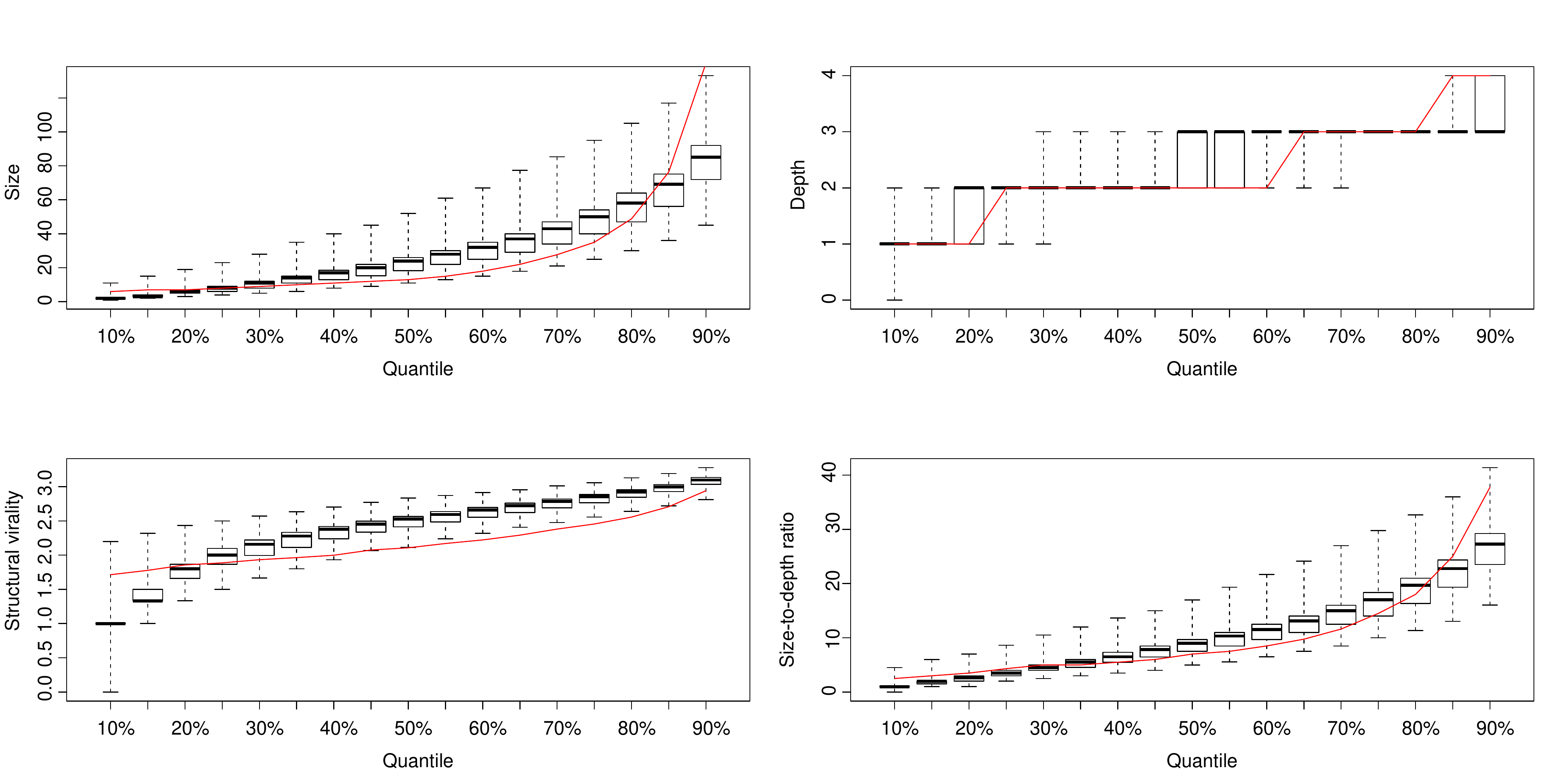}
\caption{Posterior predictive checks comparing the observed retweet cascades (red) against simulated ones (black) across different statistics describing structural properties of online spreading.}
\label{fig:posterior_checks}
\end{figure}

The results are given in Fig.~\ref{fig:posterior_checks}. For the structural properties of interest, the figure compares the distribution of the observed data against the simulated retweet cascades. In sum, our \model model appears effective: it is capable of replicating both simple and complex cascades and generally models the overall variability in the data to a large extent. This hold especially true for two statistics, namely (1)~the size and (2)~the depth of the cascades. Regarding (3)~structural virality, we find that the simulated cascades tend to be less viral. However, considering (4)~the size-to-depth ratio as an alternative measure of cascade complexity, we again find that the complexity of the data is captured adequately. 

\section{Robustness Checks}
\label{appendix:robustness}

The \model can be parameterized with arbitrary memory kernels that differ in how a tweet modulates the intensity of the self-exciting process. When changing the memory kernels (see Tbl.~\ref{tbl:model_comparison}), the prediction performance remains robust. 

\begin{table}[H]
\centering
\sisetup{table-format=6.1,group-separator={},parse-numbers=true}
{\scriptsize
\begin{tabular}{ll SS SS S[table-format=2.1,round-precision=1,round-mode=places]}
\toprule
\multicolumn{2}{c}{Kernel}& \multicolumn{2}{c}{elpd} & \multicolumn{2}{c}{lpd} & \\\cmidrule(lr){1-2}
\cmidrule(lr){3-4} \cmidrule(lr){5-6} 
Root &Non-Root & \multicolumn{1}{c}{LOOIC} & \multicolumn{1}{c}{WAIC} &
\multicolumn{1}{c}{In-sample} & \multicolumn{1}{c}{Out-of-sample} &{AUC}\\
\midrule
Power & Weibull & 251660.4 & 254705.0 & 245514.2 & 2149127& 69.44875\\
Power & Exponential & 300342.3 &303626.0& 294247.6& 2552881& 69.36406\\
Power & Power & 192734.3& 195777.6& 187409.2 &1655922& 67.66800\\
Weibull & Weibull&252209.4 &255238.3& 246293.8 &2174817 &69.42678\\
Weibull & Exponential&301212.6& 304172.2& 294871.9& 2577902& 69.32153\\
Weibull & Power&196195.7& 199071.6& 190570.1& 1707475& 67.63749\\
Exponential & Weibull&340991.3& 344272.0& 334821.3& 2982716& 69.44054\\
Exponential & Exponential& 389914.8& 393210.8& 383382.3& 3385714& 69.25874\\
Exponential & Power &      284969.9& 287985.1& 279348.5& 2517278& 67.59128\\
\bottomrule	
\multicolumn{7}{p{0.95\columnwidth}}{Metrics for model selection \citep{Gelman.2014}: log predictive density (lpd) and expected lpd (elpd), \ie, leave-one-out information criterion (LOOIC) and widely applicable information criterion (WAIC)}
\end{tabular}
}
\footnotesize
\caption{Robustness checks comparing the \model across different kernels.}
\label{tbl:model_comparison}
\end{table}

\section{Details on Baseline Models}
\label{appendix:baselines}

For feature engineering, the underlying models are both linear (logistic regression) and from non-linear machine learning (random forest, gradient boosted machine, na{\"i}ve Bayes classifier, and a neural network). All of the baselines entail hyperparameters that were trained using ten-fold cross-validation for hyperparameter tuning. Hyperparameters are in Tbl.~\ref{tbl:ml_model_tuning}.

\begin{table}[H]
\centering
{\scriptsize
\begin{tabular}{l ll}
\toprule
{Model} & {Hyperparameters} & Search grid \\
\midrule
\multicolumn{3}{l}{Sequence learning}\\
\midrule
ts-SVM \citep{Ma.2015} & Cost penalty & 0, 0.25, 0.5, 1, 2, 4, 8 \\
HMM \citep{Vosoughi.2015,Vosoughi.2017}& Number of states & 1,2,4,6,8,10,12,14 \\
GRU \citep{Ma.2016} and LSTM \citep{Ma.2016}&Number latent units&8,16,32\\
&Size of hidden units & 8,16,24 \\
Fused GRU-CNN \citep{Liu.2018}&Number of latent units & 4,8,16,32  \\
&Number of output channels & 5,10,20\\
&Kernel size&3,5,7\\
&Size of hidden units & 8,16,32 \\
\midrule
\multicolumn{3}{l}{Feature engineering}\\
\midrule
Logistic regression & None & --- \\
Random forest & \#Predictors per split & 1, 2, 3, 4, 5, 6 \\
Gradient boosted machine & Number of trees & 100, 500, 1000 \\
& Learning rate & 0.001, 0.01, 0.1\\
& Interaction depth &1, 2, 3 \\
& Min obs. per leaf &1, 5, 10 \\
Na{\"i}ve Bayes & Laplace smoothing & 0, 1, 2, 3, 4, 5 \\
& Use of kernels & No, Yes \\
& Bandwidth adjustment & 0, 0.5, 1, 1.5, 2, 2.5, 3, 3.5, 4, 4.5, 5 \\
Neural network & Size of hidden layer & 1, 3, 5, 7, 9, 11 \\
& Dropout ratio & 0.0,  0.14,  0.28,  0.42,  0.56, 0.7\\
\bottomrule	
\end{tabular}
}
\caption{Hyperparameters for the benchmark models.}
\label{tbl:ml_model_tuning}
\end{table}

\end{document}